\documentclass[12pt]{article}
\usepackage{mathptmx}
\usepackage{amssymb,amsmath,amsthm,amstext}

\newtheorem{fact}{Fact}
\newtheorem{remark}{Remark}
\newtheorem{theorem}{Theorem}[section]
\newtheorem{corollary}[theorem]{Corollary}
\newtheorem{lemma}[theorem]{Lemma}
\newtheorem{claim}[theorem]{Claim}
\newtheorem{proposition}[theorem]{Proposition}
\newtheorem{property}[theorem]{Property}

\newtheorem{definition}[theorem]{Definition}
\newtheorem{observation}[theorem]{Observation}
\newtheorem{example}[theorem]{Example}

\usepackage{epsfig}
\usepackage{amsmath}

\usepackage{amsfonts}

\def\nottoobig#1{{\hbox{$\left#1\vcenter to1.111\ht\strutbox{}\right.\n@space$}}}

\if01

\newtheorem{theorem}{Theorem}[section]

\newtheorem{lemma}[theorem]{Lemma}
\newtheorem{claim}[theorem]{Claim}

\newtheorem{definition}[theorem]{Definition}

\fi

\newcommand{\bn}{B^{=n}}
\newcommand{\an}{A^{=n}}
\newcommand{\cd}{\rm{CD}}
\newcommand{\cnd}{\rm{CND}}

\newcommand{\e}{\rm{E}}

\newcommand{\singlespacing}{\let\CS=
\@currsize\renewcommand{\baselinestretch}{1}\tiny\CS}
\newcommand{\singlespacingplus}{\let\CS=
\@currsize\renewcommand{\baselinestretch}{1.25}\tiny\CS}
\newcommand{\doublespacing}{\let\CS=
\@currsize\renewcommand{\baselinestretch}{1.75}\tiny\CS}
\newcommand{\draftspacing}{\let\CS=
\@currsize\renewcommand{\baselinestretch}{2.0}\tiny\CS}

\def\zo{\{0,1\}}
\def\zostar{\zo^{*}}

\def\mapping{\rightarrow}

\newcommand{\prob}{\rm Prob}

 \newcommand{\ie}{{i.e.}}
\newcommand{\poly}{\rm{poly}}

\newcommand{\p}{\rm P}

\makeatletter
\def\@listI{\leftmargin\leftmargini \parsep 4.5pt plus 1pt minus 1pt\topsep6pt plus 2pt minus 2pt \itemsep  2pt plus 2pt minus 1pt}

\let\@listi\@listI
\@listi
\makeatother

\author{{N.~V.~Vinodchandran\/}
\thanks{Department of Computer Science and Engineering, University of Nebraska-Lincoln. Part of the work was done while at Dept. of Computer Science, Johns Hopkins University. This work is supported in part by NSF grant CCF 0916525.}
\and{Marius Zimand\/}
\thanks{  Department of Computer and Information Sciences, Towson University,
Baltimore, MD.; email: mzimand@towson.edu; http://triton.towson.edu/\~{ }mzimand.
This work is supported in part 
by NSF grant CCF 1016158.}}

\begin{document}
\title{On optimal language compression for sets in PSPACE/poly\footnote{
Previous versions of this work have been presented at FCT'2011 and CCR'2012.}}

\date{}

\maketitle

\begin{abstract} We show that if ${\rm DTIME}[2^{O(n)}]$ is not included in ${\rm DSPACE}[2^{o(n)}]$, then, for every set $B$ in PSPACE/poly, all strings $x$ in $B$ of length $n$ can be represented by a string $compressed(x)$ of length at most $\log (|\bn|) + O(\log n)$, such that a polynomial-time algorithm, given $compressed(x)$, can distinguish $x$ from all the other strings in $\bn$. 
Modulo the $O(\log n)$ additive term, this achieves the information-theoretic optimum for string compression.  We also observe that optimal compression is not possible  for sets more complex than PSPACE/poly because  for any time-constructible superpolynomial function $t$, there is a set $A$ computable in space $t(n)$  such that at least one string $x$  of length $n$ requires   $compressed(x)$ to be  of length  $2\log (|\an|) $.
\end{abstract}

\smallskip

\section{Introduction}

In many practical and theoretical applications in computer science, it is important to represent information in a compressed way. If an application handles strings $x$ from a finite set $B$, it is desirable to represent every $x$ by another shorter string $compressed(x)$ such that $compressed(x)$ describes unambigously the initial $x$. Minimizing the length of $compressed(x)$ is one of the main goals of this task and ideally one would like to achieve the information-theoretic bound $|compressed(x)| \leq \log (|B|)$, for all $x \in B$. If a set $B$ is computably enumerable, a fundamental result in Kolmogorov complexity states that for all $x \in \bn$, $C(x) \leq \log(|\bn|) + O(\log n)$, where $C(x)$ is the Kolmogorov complexity of $x$, \ie, the shortest effective description of $x$ ($\bn$ is the set of strings of length $n$ in $B$). The result holds because $x$ can be described by its rank in the enumeration of $\bn$. However, in many applications, it is desirable that the unambiguous description is not merely effective, but also efficient. This leads to the general idea of considering time-bounded versions of Kolmogorov complexity. An interesting line of 
research~\cite{sip:c:randomness,bfl:j:boundedkolmogorov,blm:j:compression,lr:j:symmetry} in time-bounded Kolmogorov complexity, which we also pursue in this paper, focuses on a notion called {\em time-bounded distinguishing Kolmogorov complexity}, $\cd^t(\cdot)$, introduced by Sipser~\cite{sip:c:randomness}. We say that a description (called a  {\em program} in the  Kolmogorov complexity literature) $p$ {\em distinguishes} a string $x$ if $p$ accepts $x$ and only $x$. $\cd^{t,A}(x)$ is the size of the smallest program that distinguishes $x$ and that runs in time $t(|x|)$ with access to the oracle $A$. Sipser showed that, for every set $B$ and every length $n$, there is a string $w$  of length $\poly(n)$ such that, for every $x \in \bn$, $\cd^{poly, \bn}(x \mid w) \leq log (|\bn|) + \log \log(|\bn|) +O(1)$.  Hence  the optimal upper bound of $\log(|\bn|)$ can be achieved (almost) if we allow the distinguishing program to use an advice of polynomial length. 
Later, Buhrman, Fortnow, and Laplante~\cite{bfl:j:boundedkolmogorov} showed how to avoid the advice at the expense of the length of compression. More precisely, they showed 
 that for some polynomial $p$, for every set $B$, and every string $x \in \bn$, $\cd^{p, \bn}(x) \leq 2 \log(|\bn|) + O(\log n)$. Hence a significant question is whether the factor 2 in the length of compression is  necessary if no advice is given.   Interestingly,  Buhrman, Laplante, and Miltersen~\cite{blm:c:compression} showed that in the general setting, there are sets for which the answer is yes.
 In particular, they showed that there is a set $B$ such that for sufficiently large $n$, there exists $x\in B^{=n}$ for which $\cd^{t,A} \geq 2\log {|B^{=n}|} - 
O(1)$ for a $t$ that is super-polynomial.  

There are results in the literature  where the upper bound is $\log(|\bn|)$ ($+$ a small ``precision" term) at the price of weakening other parameters. For example,  Buhrman, Fortnow, and Laplante~\cite{bfl:j:boundedkolmogorov} showed such  an upper bound of $\log(|\bn|)$ in an average-case setting: For any $B$, any $\epsilon$, for all except a fraction of $\epsilon$ strings $x \in \bn$, $\cd^{poly, \bn}(x ) \leq log (|\bn|) + \poly\log (n \cdot 1/\epsilon)$. The precision term  $\poly\log (n \cdot 1/\epsilon)$ has been improved to $O(\log (n \cdot 1/\epsilon))$ in~\cite{bmvz:t:shortlist}.  In addition, Buhrman, Lee, and van Melkebeek~\cite{blm:j:compression} showed that for all $B$ and $x \in \bn$, $\cnd^{poly, \bn}(x ) \leq log (|\bn|) + O((\sqrt{\log(|\bn|)} + \log n) \log n)$, where $\cnd$ is similar to $\cd$ except that the distinguisher program is nondeterministic.
\smallskip

\subsubsection*{Our Contribution}

In this paper we consider the following question. Is it possible to achieve the optimal $\log |\bn|$ bound for the language compression problem in the case where we have a bound on the complexity of the language we are compressing?  Our main result is that, under a certain reasonable hardness assumption that is used in the area of derandomization, the upper bound of $\log (|\bn|)$ holds for every set $B$ in the class {\rm PSPACE/poly}. We state the result below. For a precise and stronger statement, see Theorem~\ref{t:main}.

\vspace{1cm}

\noindent\emph{Main Result.} {\em
Assume that there exists $f \in \e$ that cannot be computed in space $2^{o(n)}$. Then  for any $A$ in {\rm PSPACE/poly}, there exists a polynomial $p$ such that for every $x \in \an$, 
\[
\cd^{p, \an}(x) \leq \log(|\an|) + O(\log n). 
\]
}

The main result is a corollary of the following stronger result: Under the same hardness assumption, the distinguisher program $p$ for $x$ of length $\log( |\an|) + O(\log n)$ is simple conditioned by $x$, in the sense that $C^{\poly}(p \mid x) = O(\log n)$, where $C^{\poly}(\cdot)$ is the polynomial-time bounded Kolmogorov complexity function. The idea of the proof is that the hardness assumption implies efficient pseudo-random generators that are sufficient to {\em derandomize} Sipser's non-uniform result (which is a consequence of a randomized construction). 

We also show that in a natural sense the main result is optimal by showing that for any super-polynomial function $t$ that is time constructible, there is a set $A$ computable in space $t(n)$ for which the lower bound from~\cite{blm:c:compression} holds: For every polynomial $p$, for every sufficiently large $n$, there exists $x \in \an$ with $\cd^{p, \an}(x)  \geq 2\log(|\an|) - O(1)$.

Finally we consider the applicability of pseudo-random generator constructions for the problem of   (efficiently) compressing/decompressing efficient languages, originally considered by Goldberg and Sipser~\cite{gol-sip:c:compression} and subsequently treated by Trevisan, Vadhan, and Zuckerman~\cite{tre-vad-zuc:j:compression}. We show that  the hardness assumption that there exists $f \in \e$ that cannot be computed in space $2^{o(n)}$, leads to improved results on this compression problem (see Section~\ref{sec:standard} for precise statement of results.) 

It is natural to investigate to what extent basic results in Kolmogorov complexity
remain valid in resource-bounded Kolmogorov complexity, especially for the case of
polynomial bounds. In general, proofs in Kolmogorov complexity can be easily adapted for the  \emph{space-bounded} version. The case of \emph{time-bounded} Kolmogorov complexity is quite
different. Most proofs cannot be converted from the classical setting to the polynomial-time setting and, in fact, some results no longer hold. For example, symmetry of information does not hold for polynomial-time-bounded Kolmogorov complexity~\cite{lon-wat:j:symmetry} (provided one-way permutations exist, which is generally
believed to be true).

Recently, certain techniques based on the theory of pseudo-randomness in computational complexity have been used to obtain interesting results in polynomial-time-bounded Kolmogorov complexity. For instance, Antunes and Fortnow~\cite{ant-for:c:polyunivmeasure} have shown that, under a certain reasonable hardness assumption, $2^{-C^p(x)}$ dominates all P-samplable distributions (where $C^p(x)$ is the length of the shortest program that generates $x$ within time $p(|x|)$). Our approach uses the similar machinery and reinforces the fact that this is a powerful technique in time-bounded Kolmogorov complexity research.

\section{Preliminaries}

We use $|A|$ to denote the size of a finite set $A$, $[M]$ to denote the finite set $\{1,\ldots, M\}$, and  $\log(.)$ to denote $\log_2(.)$. 

We use standard notation and concepts from computational complexity and Kolmogorov complexity (see~\cite{aro-bar:b:complexity},~\cite{li-vit:b:kolmbook}). We are interested in languages in {\rm PSPACE/poly}. A language $A$ is in {\rm PSPACE/poly} if there exist a machine $M$ and  a polynomial $p$ such that for all input lengths $n$ there is a string $y$ of length $p(n)$, and for all $x$ of length $n$, $x\in A$ if and only if $M(x,y)$ accepts $x$ in $p(n)$ space. The advice represents information that is nonuniform across input lengths, and is given to the machine along with the input.

$C^{t}(x)$ denotes the $t$-time-bounded Kolmogorov complexity of $x$. Formally,  
$C^t(x)$ is the minimal  length of a program $p$ such that a fixed universal machine $U$ on input $p$ prints $x$ in at most $t(|x|)$ steps. Since a different universal machine affects the Kolmogorov complexity only by an $O(1)$ additive term, and the time bound by a $O(\log t)$ multiplicative factor, in the rest of the paper we will fix one arbitrary universal machine and work with it.

Our main focus in this paper is a variation of time-bounded Kolmogorov complexity called {\em distinguishing complexity} introduced by Sipser~\cite{sip:c:randomness}. 
The $t$-time bounded distinguishing complexity of $x$, denoted $\cd^t(x )$ 
is the length of the shortest program $p$ that accepts $x$ and only $x$ within time $t(|x|)$. We define this formally next. 

\begin{definition}[{Distinguishing complexity}]
The $t$-time bounded distinguishing complexity of $x$, denoted $\cd^t(x)$, 
is the length of the shortest program $p$ such that (1) $U(p,x)$ accepts, (2) $U(p,v)$ rejects for all $v \not= x$, and  (3) $U(p,v)$ halts in at most $t(|v|)$ steps for all $v$.
\end{definition}

Here $U$ is the type of universal Turing machine typically used for time-bounded Kolmogorov complexity. If $U$ is an oracle machine, we define in a similar way $\cd^{t,A}(x)$, by allowing $U$ to query the oracle $A$. We fix $U$ and we call a string $p$ as above, a program. We use the notation $p(x)$ as a substitute for $U(p,x)$ and $p^A(x)$ as a  substitute for $U^A(p,x)$ (\ie, $A$ is the oracle used by program $p$).

\if
 Buhrman, Fortnow, and Laplante~\cite{bfl:j:boundedkolmogorov} show that for some polynomial $p$, for every set $A$, and every string $x \in \bn$, $\cd^{p, \an}(x) \leq 2 \log(|\an|) + O(\log n)$.
This is not optimal compression because of the factor $2$, but Buhrman, Laplante, and Miltersen~\cite{blm:c:compression} show that, for some sets $A$, the factor $2$ is necessary. 

There are some results where the upper bound is asymptotically $\log(|\bn|)$ at the price of weakening other parameters.
Sipser~\cite{sip:c:randomness} shows that the upper bound of $\log(|\bn|)$ can be achieved if we allow the distinguisher program to use polynomial advice: For every set $B$, there is a string $w_B$ of length $\poly(n)$ such that for every $x \in \bn$, $\cd^{poly, \bn}(x \mid w_B) \leq log (|\bn|) + \log \log(|\bn|) +O(1)$.

As mentioned, for sets in P, NP, PSPACE, optimal compression can be achieved, using some reasonable hardness assumptions~\cite{zim:c:pspacecompression}. We show that this also holds for sets in P/poly, \ie, for sets computable by polynomial-size circuits. Furthermore, we sketch here a different proof method (even though some key elements are common), which needs a weaker hardness assumption.
\fi 

Our main tool is hardness based pseudo-random generators which we discuss next.  

\subsubsection*{Hardness assumptions and pseudo-random generators}

The proof of the main result relies on probability distributions that (1) have small support,  (2) are efficiently samplable,  and (3) cannot be distinguished from the uniform distribution by certain predicates. More precisely, we need pseudo-random generators that extend a seed of length $O(\log n)$ to a string of length $n$ in time polynomial in $n$, and such that the output ``looks" uniformly random to certain predicates $T$ of bounded complexity. Formally, a pseudo-random generator $g: \zo^{c \log n} \mapping \zo^n$ fools a predicate $T$ if
\[
| \prob_{z \in \zo^{c \log n}} [T(g(s))=1] - \prob _{z \in \zo^n}[T(u)=1] | < 1/n.
\]

By the celebrated result of Impagliazzo and Wigderson~~\cite{imp-wig:c:pbpp}, strengthening  an earlier result of Nisan and Wigderson~\cite{nis-wig:j:hard}, and its relativization obtained by Klivans and van Melkebeek~\cite{km:j:prgenoracle}, it follows that certain hardness assumptions imply the existence of pseudo-random generators of the type that we need. Let $f:\zostar \mapping \zo$ be some function and $T \subseteq \zostar$ be  a set (viewed also as a predicate via the identification with its characteristic function). Let $S_f^{T}(n)$ denote the size of the smallest circuit with $T$ gates that computes the function $f$ for inputs of length $n$.  For ${\cal C}$  a complexity class (such as PSPACE, NP,  or $\Sigma_p^k$, the $k$-th level of the polynomial hierarchy), we use $S_f^{\cal C}(n)$  to denote  $S_f^{T}(n)$ for some predicate $T$ that is complete under polynomial-time reduction for the class ${\cal C}$.   We denote 
$\e = \cup_{c > 0} {\rm DTIME}[2^{cn}]$.
 \medskip

\if{
\fbox{
Assumption $H(T)$:
There exists a function $f$ in $\e$ such that, for some $\epsilon > 0$, $S_f^{T}(n) > 2^{\epsilon n}$.
}
}
\fi

\noindent
\underline{{\em Assumption $H(T)$}}:
There exists a function $f$ in $\e$ such that, for some $\epsilon > 0$, $S_f^{T}(n) > 2^{\epsilon n}$.

\medskip
 
\begin{theorem}[Klivans and van Melkebeek~\cite{km:j:prgenoracle}] If $H(T)$ is true, then there exists a constant $c$ and a pseudo-random generator $g: \zo^{c \log n} \mapping \zo^n$ that fools the predicate $T$  and such that, for every $s \in \zo^{c \log n}$, $g(s)$ is computable in time polynomial in $n$.
\end{theorem}
In our main application the set $T$ is in PSPACE/poly. For such $T$, one can use the following hardness assumption $H_1$  that is less technical and is still plausible.
 \medskip

\noindent
\underline{{\em Assumption $H_1$}}:
There exists a function $f$ in $\e$ such that, for some $\epsilon > 0$, $S_f^{{\rm PSPACE}}(n) > 2^{\epsilon n}$.

\if{
\fbox{
Assumption $H_1$:
There exists a function $f$ in $\e$ such that, for some $\epsilon > 0$, $S_f^{{\rm PSPACE}}(n) > 2^{\epsilon n}$.
}}
\fi

\medskip


The following lemma is easy to prove.
\begin{lemma}
If $T \in {\rm PSPACE}/\poly$, then $H_1$ implies $H(T)$.
\end{lemma}
One can also use a hardness assumption that only involves uniform computation which is more cleaner to state (i.e., no circuits or advice information).
\medskip

\if{
\fbox{
Assumption $H_2$:
There exists a function $f$ in $\e$ which is not computable in space $2^{o(n)}$. 
}}
\fi

\noindent\underline{{\em Assumption $H_2$}}:
There exists a function $f$ in $\e$ which is not computable in space $2^{o(n)}$.

\medskip

 More explictly, this means that $f$ is in $\e$, and for every machine $M$ that computes $f$ there exists a constant $\epsilon > 0$ such that, for all sufficiently large $n$, $M$ requires space at least $2^{\epsilon n}$, on some input of length $n$.
\begin{lemma}[Miltersen~\cite{mil:b:derandsurvey}]
$H_2$ implies $H_1$.
\end{lemma}

\section{Main result}

In this section we state and prove our main theorem. 

\begin{theorem}
\label{t:main}
Assume $H_1$ holds. Then for every $A$ in {\rm PSPACE/poly}, there exists a polynomial $p$ such that, for all $x \in \an$, $\cd^{p,\an}(x) \leq \log|\an| + O(\log n)$.

\end{theorem}
\proof  Let $A$ in PSPACE/poly. Fix $n$, and let $k = \lceil \log |\an| \rceil$.  Let ${\cal H}$ be the set of linear functions $h:\zo^n \mapping \zo^{k+1}$ . Each $h \in {\cal H}$ is given by a  $(k+1) \times n$ matrix $H$ over GF[$2$] and $h(x) = Hx$ . We say that ``$h$ isolates $x$'' if for all $y \in \an \setminus \{x\}$, $h(x) \not= h(y)$.  It is easy to check that,  for fixed $x$ and $y$ in $\an$, $\prob_{h \in {\cal H}} [h(x) = h(y)] = 1/2^{k+1}$. Thus, for fixed $x$ in $\an$, $\prob_{h \in {\cal H}}[h \mbox{ does not isolate } x] \leq |A| \cdot 1/2^{k+1} \leq 1/2$. If we take uniformly at random a tuplet of $(k+1)$ functions $\overline{h} = (h_1, \ldots, h_{k+1}) \in {\cal H}^{k+1}$, 
$\prob_{\overline{h}}[\mbox{no } h_i \in \overline{h} \mbox{ isolates } x] \leq 1/2^{k+1}$. Therefore,
$\prob_{\overline{h}}[(\exists x \in \an) \mbox{ no } h_i \in \overline{h} \mbox{ isolates } x] \leq |A| \cdot  1/2^{k+1} \leq 1/2$.

Note that given  $\overline{h} = (h_1, \ldots,  h_{k+1})$ such that some $h_i \in \overline{h}$ isolates $x$, $x$ can be described (in the sense of $\cd^{\poly, A}()$) by $i$ and $h_i(x)$, an information which has length $\log k + (k+1) = \log |\an| + O(\log n)$. The problem is that the length of $\overline{h}$ is $n(k+1)^2$.

We can obtain a shorter such $\overline{h}$ using the assumption $H_1$ and the pseudorandom generator implied by it. Let $T(\overline{h})$, where $\overline{h} = (h_1, \ldots, h_{k+1}) \in {\cal H}^{k+1}$,  be the predicate defined by 
\begin{equation}
\label{e:t}
T(\overline{h})=1 \mbox{  iff }  (\forall x \in \an)(\exists h_i \in \overline{h}, h_i \mbox{ isolates } x).
\end{equation}
Since $A$ is in PSPACE/poly, it is easy to check that the predicate $T$ is computable in PSPACE/poly. Therefore assumption $H_1$ implies assumption $H(T)$, which, at its turn, implies the existence of a pseudo-random generator $g:\zo^{c \log n} \mapping \zo^{n (k+1)^2}$ such that
\[
\prob_{s \in \zo^{c \log n}} [T(g(s)) = 1] \geq \prob_{\overline{h}} [ T(\overline{h})=1] - 1/n \geq 1/2 - 1/n > 0.
\]
Therefore, there exists $s \in \zo^{c \log n}$ such that $g(s)$ is a tuplet $\overline{h}= (h_1, \ldots, h_{k+1})$ with the property that for every $x \in A$, there is some $i \in \{1, \ldots, k+1\}$ such that $h_i$ isolates $x$. Thus, any string $x$ in $A$ can be described by $p = (k, s, i, h_i(x))$, an information which can be encoded with $k+ O(\log n)$ bits. Indeed,  on input $(p,\nu)$,  the distinguishing algorithm constructs the pseudo-random generator $g$ (which depends on $k$ and $n = |x|$), calculates $g(s) = (h_1, \ldots, h_{k+1})$, and accepts if and only if $h_i(\nu) = h_i(x)$. By the discussion above, this algorithm only accepts the string $x$.~\qed
\smallskip

\noindent
\textbf{Remark.}~The proof shows more:  The program $p = (k, s,i, h_i(x))$ witnessing that  $\cd^{\poly, \an}(x) \leq \log | \an | + O(\log n)$, can be obtained  from $x,k,  i, s$  in polynomial time, and therefore $C^{\poly}(p \mid x) \leq O( \log n)$.
\smallskip

Using the same technique, the following result for sets in the polynomial-time hierarchy  with polynomial advice holds (the conclusion is weaker than in Theorem~\ref{t:main}, but so is the assumption).

\begin{theorem}
\label{t:general}
Assume that there exists a function $f$ in $E$ such that, for some $\epsilon > 0$, 
$S_f^{\Sigma^p_k}(n) \geq 2^{\epsilon n}$, where $k$ is a natural number. 
Then for every $A$ in $\Sigma^P_k$/{\rm poly}, there exists a polynomial $p$ such that, for all $x \in \an$, $\cd^{p,A}(x) \leq \log|\an| + O(\log n)$.

\end{theorem}

\section{A lower bound for sets computable in superpolynomial space}

We show that the complexity class PSPACE/poly is essentiallly the largest class for which the optimal language compression in Theorem~\ref{t:main} holds. Indeed, if $t(n)$ is  a function that is superpolynomial and time-constructible, then there exists a set $A$ computable in space $t(n)$ for which the $2 \log | \an | - O(1)$ lower bound shown by Buhrman, Laplante, and Miltersen~\cite{blm:c:compression} holds.

\begin{theorem} 
\label{t:lowerbound}
Let $t(n)$ be a superpolynomial and time-constructible function. There exists a set $A$ computable in space $t(n)$ such that for all sufficiently large $n$, there exists $x \in \an$ such that
\begin{itemize}
\item[(1)] $\cd^{t(n)^{1/2},A} (x) \geq 2 \log | \an | - O(1)$,
\item[(2)] $|\an| \geq t(n)^{1/6}$.
\end{itemize}

\end{theorem}
The proof follows closely the arguments from~\cite{blm:c:compression} (we add only elements that determine the space complexity bound for the set $A$). For completeness, we present it below. The key part is a combinatorial result on $k$-cover-free subsets. A $k$-cover-free family ${\cal F}$ is a family of $N$ subsets of $[M]$ with the property that no subset in ${\cal F}$ is contained in the union of $k$ other subsets in  ${\cal F}$. More precisely, 
 if $F_0, F_1, \ldots, F_k$ are distinct subsets in ${\cal F}$, then $F_0 \not\subseteq F_1 \cup \ldots \cup F_k$. Dyachkov and Rykov~\cite{dya-ryk:j:coverfree} have shown that if $k \leq N^{1/3}$, then $M \geq \frac{k^2 \log N}{2 \log k + c}$, for some constant $c$. 

\begin{theorem}[Dyachkov and Rykov~\cite{dya-ryk:j:coverfree}]
Let ${\cal F}$ be a family of $N$ subsets of $[M]$. Then if ${\cal F}$ is $k$-cover-free for 
$k \leq N^{1/3}$, then $M \geq \frac{k^2 \log N}{2 \log k + c}$, for some constant $c$. 
\end{theorem}

\noindent
{\em Proof of Theorem~\ref{t:lowerbound}.}  Let us fix $n$ (sufficiently large) and define $r = \min \{\lceil t(n)^{1/2}\rceil, 2^{n/8} \}$,  and  $k = \lceil r^{1/3} \rceil$. Let ${\cal P}$ be the set of programs of length at most $2 \log (k+1) - c_1$  (for a constant $c_1$ that will be specified later), which run in time bounded by $r$. Clearly, $|{\cal P}| \leq (k+1)^2/c_1$.

First we find the lexicographically smallest string $y$ of length $r \cdot n$ such that if $y = x_1 x_2 \ldots x_r$, with every $x_i \in \zo^n$, then the strings $x_1, \ldots, x_r$ are all distinct and $C^{3r}(y) \geq r \cdot n - 1$ (recall that $C^{3r}(y)$ is the  Kolmogorov complexity of $y$ with time bounded by $3r$). By a simple counting argument such a string exists and, furthermore, it can be found in space bounded by $r \cdot n + 3r = o(t(n)^{1/2})$.  Let $B = \{x_1, \ldots, x_r\}$.  Note that for any two distinct string $x_i, x_j \in B$, $C^r (x_i \mid x_j) \geq n/2$ (otherwise $y$ could be reconstructed in $3r$ steps from information less than $r \cdot n -1$).  

This implies that for any set $A \subseteq B$, for any program $p \in {\cal P}$, and for any $x \in A$, 
\begin{equation}
\label{e:e1}
p^{\{x\}} (x) = p^A(x). 
\end{equation}
This is true because otherwise, the program $p$  on input $x$ and with oracle $\{x\}$, during its $r$-step computation,  queries the oracle about  some string $u \in A\setminus \{x\}$.    Note that  $C^r (u \mid x) \leq \log r + |p| + O(1) < n/2$, which contradicts the property of the elements of $B$. 

\begin{claim}
\label{f:fone}
There exists a set $A \subseteq B$, $|A| = k+1$ and $x \in A$ such that for every $p \in {\cal P}$, either
\begin{itemize}
\item[(1)] $p^{\{x\}}(x)$ does not accept, or
\smallskip

\item[(2)] $p^{\{x\}}(x)$ accepts and there exists $y \in A \setminus \{x\}$ such that $p^{\{y\}}(y)$ accepts.
\end{itemize}
\end{claim}

The statement of the theorem is a consequence of  Claim~\ref{f:fone}, as can be seen from the following two observations. First, by Equation~(\ref{e:e1}), for every $p \in {\cal P}$, either
\smallskip

(1) $p^{A}(x)$ does not accept, or
\smallskip

(2) $p^{A}(x)$ accepts and there exists $y \in A \setminus \{x\}$ such that $p^{A}(y)$ accepts.
\smallskip

This implies that $\cd^{r,A}(x) \geq 2 \log(k+1) - O(1) = 2 \log |\an| - O(1)$. Secondly, it is easy to see that, by  an exhaustive search,  one can find  a set $A$ satisfying the conditions in Fact~\ref{f:fone} and print out its elements in space $O(r)$, and thus the set $A$ is computable in space $t(n)^{1/2}$.

\noindent{\em Proof of Claim~\ref{f:fone}}.
It remains to show Claim~\ref{f:fone}. To reach a contradiction,  suppose that for every subset $A \subseteq B$ of size $k+1$ and for every $x \in A$ there exists $p \in {\cal P}$ such that 
$\{u \in A \mid p^{\{u\}} (u) \mbox{ accepts } \}  = \{x\}$.  We define a family of $r$ subsets as follows: For every $x \in B$, let $F_x = \{p \in {\cal P} \mid   
 p^{\{x\}}(x) \mbox{ accepts }\}$.  $\{F_x \mid x \in B\}$  is a family of $r$ subsets of  $\{1, \ldots, (k+1)^2/c_1\}$  which by our assumption is $k$-cover-free.
By the combinatorial result of Diachkov and Rykov, $ \frac{(k+1)^2}{c_1} \geq \frac{k^2 \log r}{2 \log k + c}$, which, for large enough $c_1$,   is a contradiction.~\qed

\section{Compression of efficient languages}
\label{sec:standard}

We show that hardness based pseudo-random generator constructions can be used to get improved upper bound on standard compression/decompression problem of efficient languages first considered by  Goldberg and Sipser~\cite{gol-sip:c:compression} (this is different from the language compression problem given by the $\cd^{\poly}( \cdot)$ complexity). In the standard compression problem there are two steps: (1) given a string $x$ we seek a succinct representation of it, the string $y$ (compression), and (2) given $y$, we want to reconstruct $x$ (decompression). 

\begin{definition}
A set $A$ is compressible by a uniform family of algorithms $(Enc_n, Dec_n)_{n \geq 1}$ to length $m(n)$ if for each $n$,
\begin{itemize}
\item[(1)] $Enc_n :  \zo^n \mapping \zo^{m(n)},  Dec_n : \zo^{m(n)} \mapping \zo^n$,
\item[(2)] For all $x \in \an$, $Dec_n(Enc_n(x))= x$.
\end{itemize}
\end{definition}

Goldberg and Sipser~\cite{gol-sip:c:compression} showed that any set $A$ in $\p$ with $|\an| \leq 
2^n/n^k$, can be compressed in polynomial time to length $n - (k-3)\log n$. Their 
algorithm is probabilistic and hence can be erroneous with some small probability.  Later,
Trevisan, Vadhan and Zuckerman~\cite{tre-vad-zuc:j:compression} gave, for any set $A$ in $\p$,  
compression algorithms to length $k(n) + {\rm polylog}(n-k(n))$ that run in time 
$2^{n-k(n)} \cdot \poly(n) \cdot 2^{\poly \log (1/\epsilon)}$, where  $k(n) \geq \lceil \log |\an | \rceil$ and $k(n)$ is computable in time $\poly(n)$.  However, their algorithms 
only work for a $(1-\epsilon)$ fraction of strings in $\an$. 
We show that, under the hardness assumption $H_1$,  compression/decompression can be done for all strings in a language in P with parameters essentially identical to the above results, but without any error. 

\if{
The results in~\cite{gol-sip:c:compression} and~\cite{tre-vad-zuc:j:compression}  do not use any unproven assumption, but contain  an element of  error. 
}
\fi

\begin{theorem}
Assume $H_1$.  Let $A$ be a set in P such that for every $n$, $|\an| \leq 2^n/n^2$. 
%
%
 Let $k(n) \geq  \lceil \log |\an | \rceil$ be a function computable in polynomial time on input $1^n$. Then $A$ is compressible to length $k(n) + O(\log n)$ by algorithms $(Enc_n, Dec_n)$ running in time $2^{n-k(n)} \cdot \poly(n)$. 
\end{theorem}

\noindent
{\bf Remarks:} We make two remarks about this theorem. (1) If $k(n) = n - O(\log n)$, $(Enc_n, Dec_n)$ run in polynomial time, are deterministic,  and work for all strings in $\an$. (2) The assumption $H_1$ can be replaced by the (probably)  weaker assumption: There exists $f$ in $E$ such that,  for some $\epsilon >0$, $S_f^{\rm NP} > 2^{\epsilon n}$.

\begin{proof}
The proof is a variation of the proof of Theorem~\ref{t:main}, and we use the notation from that proof. We fix length $n$ and let $k = k(n)$.  We define the predicate $\tilde{T}$ by adding to the predicate $T$ from equation~(\ref{e:t}), the condition that all the linear functions $h_i$ have rank $(k+1)$.
Thus,
\[
\tilde{T} (\overline{h}) = 1) \mbox { iff }   T(\overline{h})= 1 \mbox { and } (\forall h_i \in \overline{h}, h_i \mbox{ has rank } k+1) ]. 
\]
The probability, when $\overline{h}$ is chosen at random in ${\cal H}^{k+1}$,  of the event ``$\tilde{T}(\overline{h}) = 1$"   is at least $1/3$, because the probability that a random $(k+1) \times n$ matrix has rank less than $k+1$ is approximately $O(2^{-(n-k)})$, and therefore the probability of the event  ``$\tilde{T} (\overline{h}) = 1$"  is only slightly less than the probability  of the event  ``$T(\overline{h})= 1$,"  which we have seen to be at least $1/2$.
Consequently, there exists a pseudo-random generator $g_k:\zo^{c \log n} \mapping \zo^{n(k+1)^2}$ such that 
\[
\prob_{s \in \zo^{c \log n}}[ \tilde{T}(g_k(s))=1] > 0.
\]
The compression algorithm $Enc_n$ on input $x \in \an$ works as follows: 
It  finds a  seed $s$ such that if $g_k(s) = (h_1, \ldots, h_{k+1})$, there exists $h_i$ that isolates $x$ and such that $h_i$ has rank $(k+1)$.
 This takes time $2^{n-k} \cdot \poly(n)$ because if $h_i$ has rank $(k+1)$, there are at most $2^{n-k-1}$ strings $x'$ such that $h_i(x') = h_i(x)$.  
Then, the algorithm $Enc_n$ on input $x$ returns $(k,n, s,i,h_i(x))$.

The decompression algorithm $Dec_n$  on input $(k,n, s,i,y)$, constructs $g_k$, then $g_k(s) = (h_1, \ldots, h_{k+1})$ and determines the set $h_i^{-1}(y)$ which has at most $2^{n-k-1}$ elements. Then it searches for the unique element in that set that belongs to $\an$, and returns that element.

\end{proof}


\end{document}